\documentclass[conference]{IEEEtran}
\makeatletter
\def\ps@headings{%
\def\@oddhead{\mbox{}\scriptsize\rightmark \hfil \thepage}%
\def\@evenhead{\scriptsize\thepage \hfil \leftmark\mbox{}}%
\def\@oddfoot{}%
\def\@evenfoot{}}
\makeatother
\usepackage{glossaries}
\usepackage{lipsum}
\usepackage{setspace}
\usepackage{bm}
\usepackage{bbm}
\usepackage{cite}
\usepackage{lettrine}
\usepackage{amssymb}
\usepackage{amsmath}
\usepackage{amsfonts}
\usepackage{enumerate}
\usepackage{enumitem}
\usepackage{indentfirst}
\usepackage{graphicx}
\usepackage{epstopdf}
\usepackage{multirow} 
\usepackage{xcolor}
\usepackage{listings}
\usepackage{float}
\usepackage{subfigure}
\usepackage{amsthm}
\usepackage[top=0.75in, bottom=1in, left=0.65in, right=0.65in]{geometry}
\usepackage{algpseudocode}
\usepackage{algorithmicx,algorithm}

\usepackage{algorithm} 
\usepackage{url}
\usepackage[compact]{titlesec} 

\newcommand{\beq}{\begin{equation}}
\newcommand{\eeq}{\end{equation}}

\newtheorem{proposition}{\bf Proposition}

\makeatletter
\newcommand{\manuallabel}[2]{\def\@currentlabel{#2}\label{#1}}
\makeatother

\allowdisplaybreaks

\newcommand{\hil}{\mathrm{H}}
\newcommand{\tran}{\top}
\newcommand{\iu}{\mathbbm{i}}

\newcommand{\trace}{\operatorname{Tr}}

\newcommand{\diag}{\operatorname{diag}}
\newcommand{\dd}{\operatorname{d}}
\renewcommand{\trace}{\operatorname{tr}}
\renewcommand{\diag}{\bm{\mathrm{diag}}}

\newcommand{\los}{\mathrm{LoS}}
\newcommand{\mulpath}{\mathrm{MP}}
\newcommand{\user}{\mathrm{U}}
\newcommand{\elem}{\mathrm{E}}

\makeglossaries
\newacronym{rhs}{RHS}{reconfigurable holographic surface}

\newacronym{isac}{ISAC}{integrated sensing and communications}

\newacronym{crlb}{CRLB}{Cramer-Rao lower bound}
\newacronym{fim}{FIM}{Fisher information matrix}

\newacronym{sgd}{SGD}{stochastic gradient descent}

\newacronym{ftl}{FTL}{federated transfer learning}

\newacronym{ml}{ML}{machine learning}

\newacronym{roi}{ROI}{region of interest}

\newacronym{pos_est}{PE}{position estimator}

\newacronym{bs}{BS}{base station}

\newacronym{rf}{RF}{radio frequency}

\newacronym{em}{EM}{electromagnetic}

\newacronym{ofdm}{OFDM}{orthogonal frequency division multiplexing}

\newacronym{pos_process}{DPP}{downlink positioning process}

\newacronym{pos_protocol}{DPP}{downlink positioning protocol}

\newacronym{pos_acc}{PP}{positioning precision}

\newacronym{mmht}{MMHT}{multi-band multi-pattern holographic transmission}

\newacronym{me}{meta-element}{metamaterial element}

\newacronym{aod}{AoD}{angle of departure}
\newacronym{aoa}{AoA}{angle of arrival}

\newacronym{los}{LoS}{line-of-sight}

\newacronym{wss}{WSS}{wide-sense stable}

\newacronym{mse}{MSE}{mean square error}

\newacronym{rss}{RSS}{received signal strength}

\newacronym{bcd}{BCD}{block coordinate descent}

\newacronym{gps}{GPS}{Global Position System}

\newacronym{uav}{UAV}{unmanned aerial vehicle}

\newacronym{psgd}{PSGD}{proximal stochastic gradient descent}

\newacronym{rfid}{RFID}{radio frequency identification}

\newacronym{snr}{SNR}{signal-to-noise ratio}

\newacronym{ris}{RIS}{reconfigurable intelligent surface}

\newacronym{mb}{MB}{multi-band}

\newacronym{tnn}{TNN}{transformer neural network}

\newacronym{and}{A\&D}{analog and digital}

\newacronym{mimo}{MIMO}{multiple-input multiple-output}

\newacronym{uwb}{UWB}{ultra-wide-band}

\newacronym{csi}{CSI}{channel state information}
\newacronym{tx}{Tx}{transmitting}
\newacronym{rx}{Rx}{receiving}
\newacronym{tdd}{TDD}{time-division duplex}
\newacronym{cg}{CG}{conjugate gradient}

\newglossaryentry{SysName}
{
    name={\acrshort{mb} \acrshort{rhs}-based \acrshort{isac} system},
    description={the name of the proposed system}
}

\newglossaryentry{MetaElem}
{
    name={metamaterial element},
    description={Metamaterial element of RHS}
}

\newglossaryentry{HoloMat}
{
    name={holographic pattern matrix},
    description={holographic pattern matrix of RHS}
}

\newglossaryentry{Protocol}
{
    name={position-then-transmit protocol},
    description={long name of the protocol}
}

\newglossaryentry{Function}
{
    name={positioner function},
    description={long name of the protocol}
}

\newglossaryentry{elemInterval}
{
    name={\ensuremath{\Delta_{\elem}}},
    sort={p},
    description={interval between adjacent units}
}

\newglossaryentry{speedLight}
{
    name={\ensuremath{v_0}},
    sort={p},
    description={speed of light}
}

\newglossaryentry{noisePowerDensity}
{
    name={\ensuremath{P_{\mathrm{N}}}},
    sort={p},
    description={spectral power density}
}

\newglossaryentry{tarSpace}
{
	name={\ensuremath{\mathcal P}},
	sort={p},
	description={region of interest}
}

\newglossaryentry{numUser}
{
	name={\ensuremath{U}},
	sort={p},
	description={number of users}
}

\newglossaryentry{numElem}
{
	name={\ensuremath{N_{\mathrm{E}}}},
	sort={p},
	description={number of metamaterial elements}
}

\newglossaryentry{numFeed}
{
    name={\ensuremath{K}},
    sort={k},
    description={number of feeds in RHS}
}

\newglossaryentry{numState}
{
    name={\ensuremath{N_{\mathrm{S}}}},
    sort={n},
    description={number of configurable states of each meta element}
}

\newglossaryentry{setRadCoeff}
{
    name={\ensuremath{\mathcal R}},
    sort={r},
    description={set of radiation coefficients}
}

\newglossaryentry{numBand}
{
	name={\ensuremath{N_{\mathrm{B}}}},
	sort={n},
	description={number of downlink ofdm bands}
}

\newglossaryentry{numSBand}
{
	name={\ensuremath{N_{\mathrm{SB}}}},
	sort={p},
	description={number of sub-band of each ofdm band}
}

\newglossaryentry{ulFreq}
{
	name={\ensuremath{f_{\mathrm{c}}^{\mathrm{u}}}},
	sort={p},
	description={carrier frequency of UL transmission}
}

\newglossaryentry{cenFreq_i}
{
	name={\ensuremath{f_{\mathrm{c},i}}},
	sort={p},
	description={center frequency}
}

\newglossaryentry{subFreq_ij}
{
	name={\ensuremath{f_{i,j}}},
	sort={p},
	description={center frequency of sub-band}
}

\newglossaryentry{subBandwidth}
{
	name={\ensuremath{W}},
	sort={p},
	description={sub-band bandwidth}
}

\newglossaryentry{ulBandwid}
{
	name={\ensuremath{B^{\mathrm{u}}}},
	sort={p},
	description={bandwidth of UL transmission}
}

\newglossaryentry{userMaxSpeed}
{
	name={\ensuremath{v_{\max}}},
	sort={p},
	description={user max speed}
}

\newglossaryentry{userPos}
{
	name={\ensuremath{\bm p^{\user}}},
	sort={p},
	description={user position}
}

\newglossaryentry{userPos_est}
{
	name={\ensuremath{\tilde{\bm p}^{\user}}},
	sort={p},
	description={estimated user position}
}

\newglossaryentry{userDistribu}
{
	name={\ensuremath{\varGamma^{\user}}},
	sort={p},
	description={distribution of user's position}
}

\newglossaryentry{posFunc}
{
    name={\ensuremath{\bm f}},
    sort={p},
    description={positioning estimator of users}
}

\newglossaryentry{numFrame}
{
    name={\ensuremath{F}},
    sort={k},
    description={number of frames sent in each band}
}

\newglossaryentry{sigMat_i}
{
    name={\ensuremath{\bm S_i}},
    sort={s},
    description={signal matrix of the $i$-th band}
}

\newglossaryentry{sigMat_ij}
{
    name={\ensuremath{\bm S_{i,j}}},
    sort={s},
    description={signal matrix of the $i$-th band}
}

\newglossaryentry{sigSet}
{
    name={\ensuremath{\{\bm S_{i,j}\}_{i,j}}},
    sort={s},
    description={set of signal matrices}
}

\newglossaryentry{sigVec_ijq}
{
    name={\ensuremath{\bm s_{i,j}^{(q)}}},
    sort={s},
    description={feed signal vector in the $i$-th band, $j$-th sub-band, and $q$-th frame}
}

\newglossaryentry{codeMat_i}
{
    name={\ensuremath{\bm C_i}},
    sort={c},
    description={code matrix of the RHS in the $i$-th band}
}

\newglossaryentry{codeSet}
{
    name={\ensuremath{\{\gls{codeMat_i}\}_i}},
    sort={c},
    description={set of code matrix of RHS}
}

\newglossaryentry{codeVec_iq}
{
    name={\ensuremath{\bm c_{i}^{(q)}}},
    sort={s},
    description={code vector of RHS in the $i$-th band and $q$-th frame}
}

\newglossaryentry{code_iqm}
{
    name={\ensuremath{c_{i,m}^{(q)}}},
    sort={s},
    description={code of Meta-elemt $m$ in the $i$-th band and $q$-th frame}
}

\newglossaryentry{recvSigMat}
{
    name={\ensuremath{\bm Y_{\mathrm{Rx}}}},
    sort={y},
    description={received signals arranged as a matrix in the all the bands}
}

\newglossaryentry{recvSigVec_i}
{
    name={\ensuremath{\bm y_{i}}},
    sort={y},
    description={received signals arranged as a vector in the Band $i$}
}

\newglossaryentry{recvSigSet}
{
    name={\ensuremath{\mathcal Y}},
    sort={y},
    description={set of all the received signal matrices}
}

\newglossaryentry{posFeed_k}
{
    name={\ensuremath{\bm p^{\user}_k}},
    sort={p},
    description={position of Feed $k$}
}

\newglossaryentry{posElem_m}
{
    name={\ensuremath{\bm p^{\mathrm{E}}_m}},
    sort={p},
    description={position of Meta-elem $m$}
}

\newglossaryentry{posElem_1}
{
    name={\ensuremath{\bm p^{\mathrm{E}}_1}},
    sort={p},
    description={position of Meta-elem $1$}
}

\newglossaryentry{holoVec_ijqm}
{
    name={\ensuremath{\tau_{i,j,m}^{(q)}}},
    sort={t},
    description={holograph record/radiation pattern in Frame $q$ of Band $i$ Sub-band $j$}
}

\newglossaryentry{recvSig_ijq}
{
    name={\ensuremath{y_{i,j}^{(q)}}},
    sort={y},
    description={received signal of user}
}

\newglossaryentry{noise_ijq}
{
	name={\ensuremath{e_{i,j}^{(q)}}},
	sort={p},
	description={thermal noise signal}
}

\newglossaryentry{losgain_ijm}
{
    name={\ensuremath{h^{\los}_{i,j,m}}},
    sort={h},
    description={LoS channel gain}
}

\newglossaryentry{mpgain_ijmq}
{
    name={\ensuremath{h^{\mulpath,(q)}_{i,j,m}}},
    sort={h},
    description={Multi-path channel gain}
}

\newglossaryentry{mpgainVec_ijq}
{
    name={\ensuremath{\bm h^{\mulpath,(q)}_{i,j}}},
    sort={h},
    description={Multi-path channel gain vector}
}

\newglossaryentry{noisepower_i}
{
    name={\ensuremath{\sigma_i^2}},
    sort={s},
    description={noise power}
}
\newglossaryentry{userMaxPower}
{
    name={\ensuremath{P_{\max}}},
    sort={s},
    description={user's max power}
}

\newglossaryentry{noisePower}
{
    name={\ensuremath{\sigma_{\mathrm{n}}^2}},
    sort={s},
    description={noise power}
}

\newglossaryentry{covmat_i}
{
    name={\ensuremath{\bm V_i}},
    sort={v},
    description={covariance matrix of multipath gain}
}

\newglossaryentry{angularResponse_i}
{
    name={\ensuremath{\bm \alpha_i}},
    sort={a},
    description={angular array response function of RHS in Band $i$}
}

\newglossaryentry{covCoefFuncFreq_i}
{
	name={\ensuremath{\rho_{\mathrm{f},i}}},
	sort={r},
	description={covariance coefficient function in the frequency domain}
}

\newglossaryentry{covCoefFuncTime_i}
{
	name={\ensuremath{\rho_{\mathrm{t},i}}},
	sort={r},
	description={covariance coefficient function in the time domain}
}

\newglossaryentry{rmsValue_i}
{
    name={\ensuremath{\sigma_{\mathrm{rms},i}}},
    sort={s},
    description={RMS delay spread value}
}

\newglossaryentry{dopplerFreq_i}
{
    name={\ensuremath{f_{\mathrm{D},i}}},
    sort={f},
    description={Doppler frequency}
}

\newglossaryentry{holoMat_i}
{
    name={\ensuremath{\bm T_i}},
    sort={t},
    description={holographic radiation pattern of RHS}
}

\newglossaryentry{bPropMat_i}
{
    name={\ensuremath{\bm B_{i}}},
    sort={b},
    description={on-board propagation coefficient matrix}
}

\newglossaryentry{gainMat_i}
{
    name={\ensuremath{\bm G_i}},
    sort={g},
    description={gain matrix in Band $i$}
}

\newglossaryentry{valueSet_i}
{
    name={\ensuremath{[1,\gls{numBand}]}},
    sort={0},
    description={shortcut forall $i$}
}

\newglossaryentry{valueSet_j}
{
    name={\ensuremath{[1,\gls{numSBand}]}},
    sort={0},
    description={shortcut forall $j$}
}

\newglossaryentry{valueSet_k}
{
    name={\ensuremath{[1,\gls{numFeed}]}},
    sort={0},
    description={shortcut forall $k$}
}

\newglossaryentry{valueSet_q}
{
    name={\ensuremath{[1,\gls{numFrame}]}},
    sort={0},
    description={shortcut forall $q$}
}

\newglossaryentry{valueSet_m}
{
    name={\ensuremath{\{1,...,\gls{numElem}\}}},
    sort={0},
    description={shortcut forall $m$}
}

\newglossaryentry{vecUserSpeed}
{
    name={\ensuremath{\bm v^{\user}}},
    sort={0},
    description={velocity vector of user}
}

\newglossaryentry{durPosProc}
{
    name={\ensuremath{D_{\mathrm{pos}}}},
    sort={0},
    description={duration of positioning process}
}

\newglossaryentry{durFrame}
{
    name={\ensuremath{\Delta_{\mathrm{t}}}},
    sort={0},
    description={duration of each frame}
}

\newglossaryentry{crlb_userpos}
{
    name={\ensuremath{\mathrm{CRLB}(\gls{userPos})}},
    sort={0},
    description={CRLB at certain user position}
}

\newglossaryentry{fimMat_userpos}
{
    name={\ensuremath{\bm I_{\mathrm{FIM}}(\gls{userPos})}},
    sort={0},
    description={FIM at certain user position}
}

\newglossaryentry{invfimMat_userpos}
{
    name={\ensuremath{\bm I^{-1}_{\mathrm{FIM}}(\gls{userPos})}},
    sort={0},
    description={FIM at certain user position}
}
\newglossaryentry{inv2fimMat_userpos}
{
    name={\ensuremath{\bm I^{-2}_{\mathrm{FIM}}(\gls{userPos})}},
    sort={0},
    description={FIM at certain user position}
}

\newglossaryentry{roi_sideLen_x}
{
    name={\ensuremath{l_{\mathrm x}}},
    sort={l},
    description={side length of ROI in x-axis}
}

\newglossaryentry{roi_sideLen_y}
{
    name={\ensuremath{l_{\mathrm y}}},
    sort={l},
    description={side length of ROI in y-axis}
}

\newglossaryentry{roi_sideLen_z}
{
    name={\ensuremath{l_{\mathrm z}}},
    sort={l},
    description={side length of ROI in z-axis}
}

\newglossaryentry{numSamples}
{
    name={\ensuremath{N_{\mathrm{sam}}}},
    sort={l},
    description={side length of ROI in z-axis}
}
\newglossaryentry{setSamples}
{
    name={\ensuremath{\mathcal S_{\mathrm{sam}}}},
    sort={l},
    description={side length of ROI in z-axis}
}

\newglossaryentry{numMaxIteration}
{
    name={\ensuremath{N_{\mathrm{itr}}}},
    sort={l},
    description={maximum number of iteration}
}

\newglossaryentry{reshapeFunc}
{
    name={\ensuremath{\bm R}},
    sort={l},
    description={reshape function}
}

\newglossaryentry{numUpdate}
{
    name={\ensuremath{N_{\mathrm{us}}}},
    sort={l},
    description={number of update steps}
}

\IEEEoverridecommandlockouts
\begin{document}
\title{\huge{Multi-band Reconfigurable Holographic Surface Based \\ISAC Systems: Design and Optimization}}

\author{
\IEEEauthorblockN{
\small{Jingzhi~Hu}\IEEEauthorrefmark{1},
\small{Zhe~Chen}\IEEEauthorrefmark{2},
\small{and~Jun~Luo}\IEEEauthorrefmark{1}\\}
\IEEEauthorblockA{
	\IEEEauthorrefmark{1}\small{School of Computer Science and Engineering, Nanyang Technological University, Singapore,}\\
	\IEEEauthorrefmark{2}\small{China-Singapore International Joint Research Institute and AIWiSe Technology Co., Ltd, Guangzhou, China.}\\}
\thanks{\copyright 2023 IEEE. Personal use of this material is permitted. Permission from IEEE must be obtained for all other uses, in any current or future media, including reprinting/ republishing this material for advertising or promotional purposes, creating new collective works, for resale or redistribution to servers or lists, or reuse of any copyrighted component of this work in other works.}
}
\maketitle

\setlength{\abovecaptionskip}{0pt}
\setlength{\belowcaptionskip}{-10pt}
\setlength{\textfloatsep}{10pt}
\begin{abstract}
Metamaterial-based reconfigurable holographic surfaces~(\acrshort{rhs}s) have been proposed as novel cost-efficient antenna arrays, which are promising for improving the positioning and communication performance of \acrfull{isac} systems.
However, due to the high frequency selectivity of the metamaterial elements, RHSs face challenges in supporting ultra-wide bandwidth~(UWB), which significantly limits the positioning precision.
In this paper, to avoid the physical limitations of UWB RHS while enhancing the performance of RHS-based ISAC systems, we propose a \acrfull{mb} \acrshort{rhs} based \acrshort{isac} system.
We analyze its positioning precision and propose an efficient algorithm to optimize the large number of variables in analog and digital beamforming.
Through comparison with benchmark results, simulation results verify the efficiency of our proposed system and algorithm, and show that the system achieves $42\%$ less positioning error, which reduces $82\%$ communication capacity loss.

\end{abstract}

\section{Introduction}
Recently, metamaterial-based reconfigurable holographic surfaces~(\acrshort{rhs}s) have been proposed as novel cost-efficient antenna arrays, which possess high potential in improving the performance of wireless networks~\cite{Deng2021Reconfigurable}.
RHSs are characterized by their large number of densely arranged metamaterial antenna elements (\emph{\acrshort{me}s}), which have smaller spatial intervals than half of their working wavelength~\cite{Boyarsky2021Electronically}.
Such dense placement enables \acrshort{rhs}s to have strong manipulation capability for \acrfull{em} waves and can synthesize various beam forms through \acrfull{and} beamforming~\cite{Zhang2022Holographic}.
By leveraging this capability, an \acrfull{bs} equipped with an \acrshort{rhs} can focus transmitted signals towards users for communication enhancement.

To achieve this enhancement, the \acrshort{rhs}-based \acrshort{bs}s need to know the \acrfull{csi} or the positions of the users.
However, the channel state estimation problem of the \acrshort{rhs}s is highly complicated due to the large number of \acrshort{me}s lacking signal processing capability~\cite{Wei2021Channel}.
Alternatively, using the \acrshort{rhs}, a \acrshort{bs} can effectively positioning a target user and focus its beam towards it. 
This indicates that the \acrshort{rhs}-based \acrshort{bs}s are intrinsically fit for \acrfull{isac} in the sense that the communication can benefit from the positioning results obtained by the \acrshort{rhs}.

In this regard, the performance of an \acrshort{rhs}-based \acrshort{isac} system is fundamentally influenced by its positioning precision.
In literature, several works have considered the positioning aspect of \acrshort{rhs}-based \acrshort{isac} systems.
In~\cite{Zhang2022Holographic}, the authors proposed a beamforming algorithm to simultaneously generate beams for communication and sensing purposes with high directional gains.
In~\cite{ZhangX2022Holographic}, the authors used \acrshort{rhs} for target detection and obtained high accuracy with low cost and power consumption.

Additionally, given the essential similarity between reconfigurable intelligent surfaces~(RISs) and \acrshort{rhs}s, the positioning techniques for \acrshort{ris}-based systems have the potential to be applied in \acrshort{rhs}-based systems\footnote{The main difference between \acrshort{ris}s and \acrshort{rhs}s is that \acrshort{ris}s perform beamforming by reflecting signals, while \acrshort{rhs}s radiate the signals themselves.}.
In~\cite{zhang2020towards}, the authors utilized an \acrshort{ris} to generate distinguishable signals strength values for different locations to facilitate positioning. 
In~\cite{Elzanaty2021Reconfigurable}, the authors proposed a \acrshort{snr}-based \acrshort{ris} configuration profile that increases the positioning precision.

However, in existing works, the fundamental limitation is that the signal bandwidths for positioning are not broad enough.
This reduces the positioning precision, as the spatial resolution of positioning typically increases with the bandwidth\cite{Sturm2009ANovel}.
Though the authors in~\cite{Ma2021Indoor} considered using \acrfull{uwb} technique in \acrshort{ris}-based positioning systems, it is challenging to implement \acrshort{ris} or \acrshort{rhs} that can support beamforming of \acrshort{uwb} signals in practice, due to the high frequency selectivity of the reconfigurable \acrshort{me}s~\cite{Deng2021Reconfigurable, Boyarsky2021Electronically}.

In this paper, we propose the \acrfull{mb} \acrshort{rhs}-based \acrshort{isac} system, where multiple uplink bands are utilized in a \acrfull{tdd} manner to enhance the positioning precision and eventually leads to better communication capacity.
The \acrshort{mb} technique is a promising alternative of \acrshort{uwb}~\cite{Jafari2015TDOA}.
As the feasibility of \acrshort{mb} \acrshort{me} has been addressed in literature~\cite{Jagadeesan2015}, the \acrshort{mb} \acrshort{rhs} becomes a promising solution to get around the bandwidth limitation and achieve high positioning precision.
To fully exploit the strength of the proposed system, we propose a \emph{\gls{Protocol}} for the system.
Based on the protocol, we establish the channel model of the system in a rich scattering environment.
Then, we derive the positioning precision of the system in terms of the \acrfull{crlb} given the \acrshort{and} beamforming of the \acrshort{rhs}.

Nevertheless, as multiple bands are adopted, the large number of controlling variables in the \acrshort{and} beamforming is further multiplied, which cannot be handled by conventional small-scale optimization algorithms efficiently. 
To handle this challenge, we derive a closed-form formula of the \acrshort{crlb}'s gradient and then propose an efficient \acrfull{bcd} algorithm where the digital and analog beamforming variables are optimized alternatingly.
Simulation results verify that with the proposed algorithm, the \acrshort{mb} \acrshort{rhs}-based system achieves lower \acrshort{crlb} in positioning, which leads to lower positioning error and much smaller communication capacity loss compared with benchmarks.

The rest of the paper is organized as follows.
In Section~\ref{sec: sys mod}, we propose the system model, the protocol, and the channel model of the \gls{SysName}.
In Section~\ref{sec: problem formulation}, the \acrshort{crlb} of positioning is derived, and the problem for beamforming optimization is formulated.
In Section~\ref{sec: alg design}, the gradients of \acrshort{crlb} are derived, and the algorithm to solve the formulated problem is designed.
Simulation results are provided in Section~\ref{sec: simulation result}, and a conclusion is drawn in Section~\ref{sec: conclu}.

\emph{Notations}: 
$\overline{(\cdot)}$, $(\cdot)^{\top}$, and $(\cdot)^{\hil}$ denote the conjugate, transpose, and Hermitian transpose. 
$\odot$ and $\otimes$ denote the Hadamard and Kronecker products. 
$\trace(\cdot)$ and $\diag(\cdot)$ return the trace and the main diagonal vector of a matrix.
$\{\bm X_i\}_i$ is the set of $\bm X_i$ for all $i$ in its value range.
$[\bm X]_{i}$ and $[\bm X]_{i,j}$ indicate the $i$-th row vector and the $(i,j)$-th element of $\bm X$.
Besides, $[\bm X]_{i:j}$ is the sub-matrix of $\bm X$ composed of its $i$-th to $j$-th row vectors.
Moreover, $\bm I_{N}$ denotes the $N$-dim unit matrix.
$\bm 1_{N}$ and $\bm J_{N}$ denote all-ones $N$-dim vector and $N\times N$ matrix, respectively.
$\Re(\cdot)$ is the real part of the argument.

\section{System Model}
\label{sec: sys mod}

In this section, we introduce the \gls{SysName}.
We describe its components, design a working protocol for it, and establish the channel model of the system.

\begin{figure}[!t]
\centerline{ \includegraphics[width=0.8\linewidth]{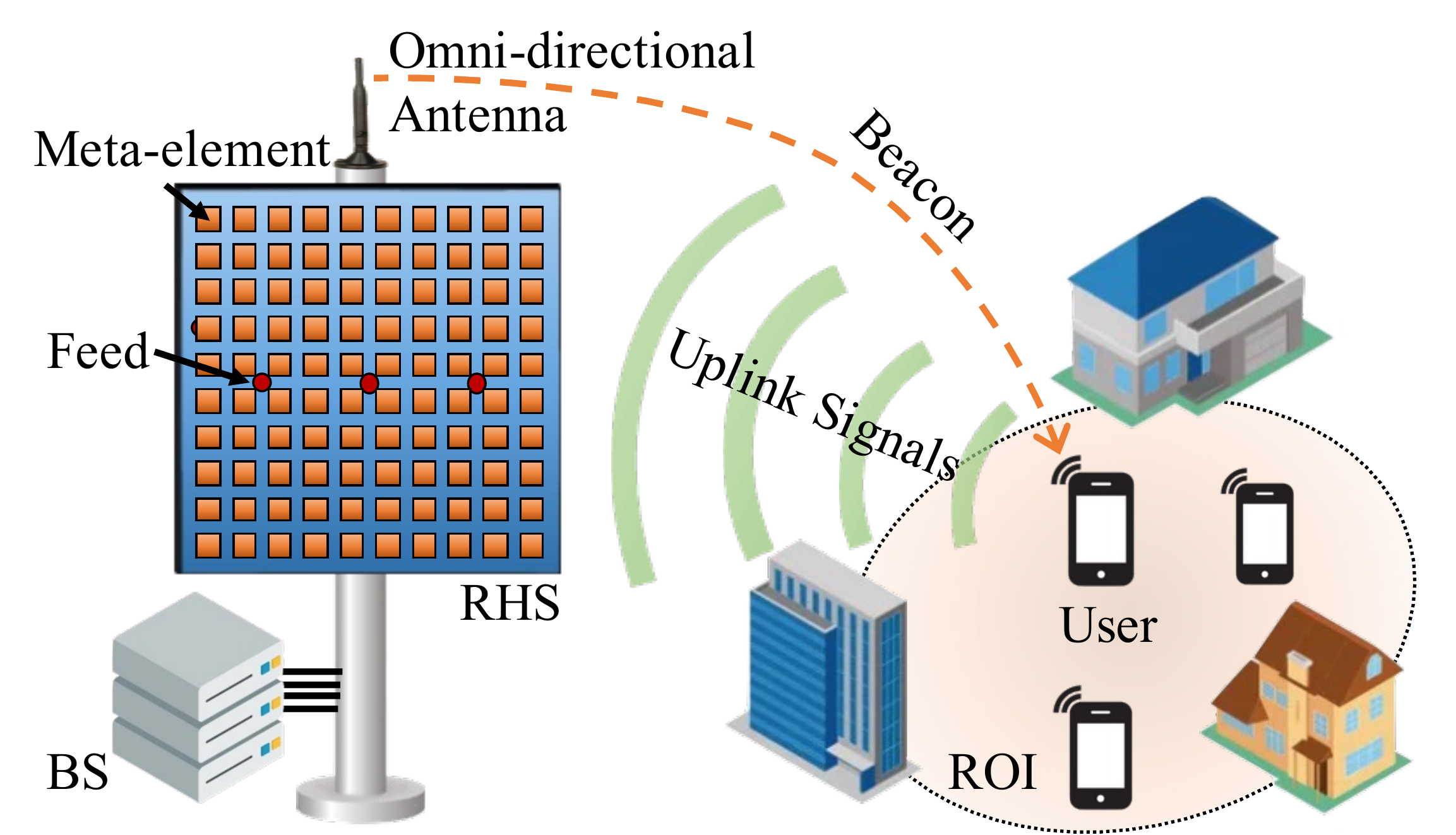} }
\caption{\gls{SysName}}
\label{fig: 1 sys mod}
\end{figure}

\subsection{System Components}
As shown in Fig.~\ref{fig: 1 sys mod}, the system contains a \acrshort{bs} equipped with a \acrshort{rhs} serving multiple users in a \acrfull{roi}.
Each user has an omni-directional antenna for \acrshort{tx} and \acrshort{rx}.
The \acrshort{bs} uses the \acrshort{rhs} for \acrfull{rx} and \acrfull{tx} \acrfull{ofdm} signals for positioning and transmitting data, while it also has an omni-directional antenna to broadcast beacons for link construction.
The \acrshort{bs} is connected to the \acrshort{rhs} via $\gls{numFeed}$ feeds, and each feed can send or receive signals from all the $\gls{numElem}$ \acrshort{me}s through on-board signal propagation.
The $\gls{numElem}$ \acrshort{me}s are arranged as a square with the interval between adjacent \acrshort{me}s being \gls{elemInterval}.
Each \acrshort{me} can be electronically configured into multiple \emph{states}, each with its own radiation coefficient.
The radiation coefficients of all the \acrshort{me}s are referred to as the \emph{configuration} of the \acrshort{rhs}, which is denoted by vector $\bm c\mathbb \in [0,1]^{1\times\gls{numElem}}$~\cite{Zhang2022Holographic}.

The \acrshort{rhs} and \acrshort{bs} are able to transmit and receive signals in $\gls{numBand}$ bands in a \acrshort{tdd} manner by adjusting their configurations.
Each band $i$~($i \in \{1,...,\gls{numBand}\}$) is centered at frequency $\gls{cenFreq_i}$ and composed of $\gls{numSBand}$ orthogonal sub-bands, each with center frequency $\gls{subFreq_ij}$~($j\in\{1,...,\gls{numSBand}\}$) and bandwidth $\gls{subBandwidth}$.
In each band, the \acrshort{bs} is able to apply \acrshort{and} beamforming. 
To be specific, the \acrshort{bs} performs the analog beamforming by controlling the configuration $\bm c$ of the \acrshort{rhs}, and performs the digital beamforming through a weighted combination of the \acrshort{tx}/\acrshort{rx} symbols for different feeds, with combining vector denoted by $\bm s\in \mathbb C^{1\times\gls{numFeed}}$.
The \acrshort{and} beamforming capability enables the \acrshort{bs} to steer \acrshort{tx}/\acrshort{rx} beams, enhancing positioning and communication.

A user positioning process is performed by the \acrshort{bs} processing the \acrshort{rx} signals from the users.
The users are within the \acrshort{roi}, which is modeled as a 3D spatial region with dimensions $\gls{roi_sideLen_x}\times\gls{roi_sideLen_y}\times \gls{roi_sideLen_z}$ m$^3$ and is rich in scatterers.
The probability distribution of user position within the \acrshort{roi} is $\gls{userDistribu}$, and the maximum speed of each user is $\gls{userMaxSpeed}$.
We assume that $\gls{userMaxSpeed}$ is not large and thus the users' positions can be considered fixed during each positioning process.

\subsection{Protocol Design}
\begin{figure}[!t]
\centerline{ \includegraphics[width=1\linewidth]{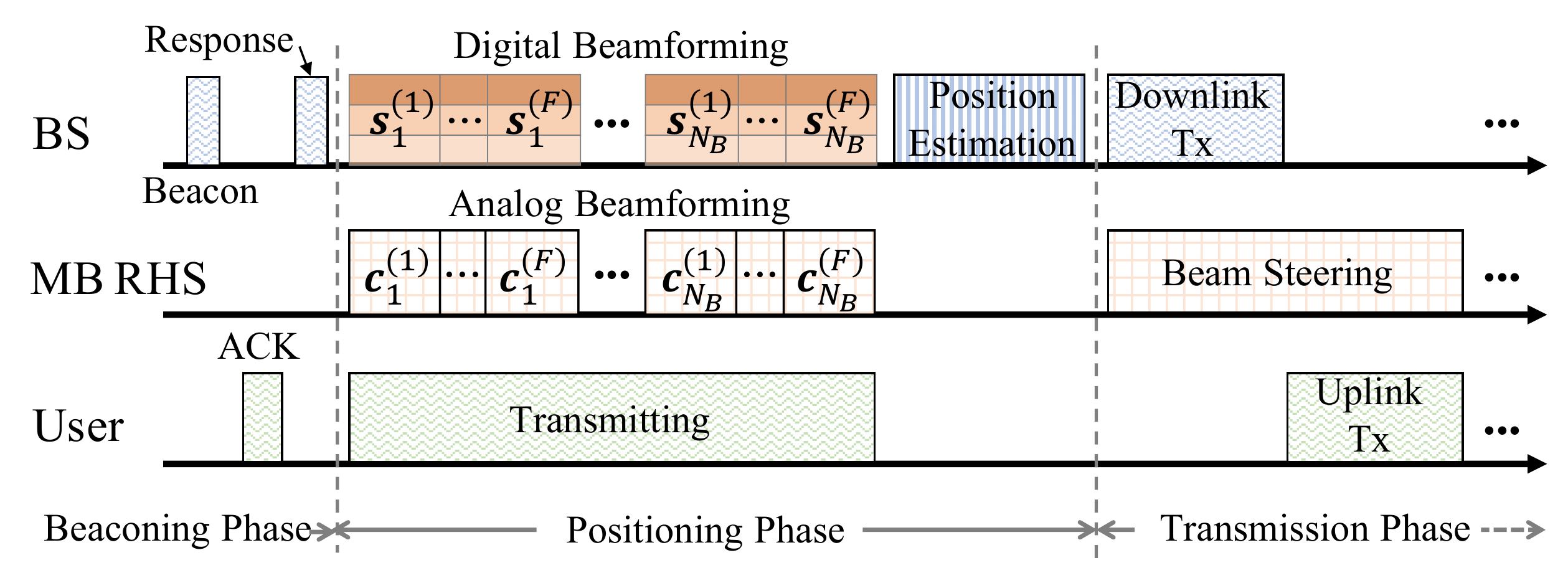} }
\vspace{-0.5em}
\caption{Position-then-transmit protocol}
\label{fig: 2 protocol}\
\vspace{-0.5em}
\end{figure}

We propose a \gls{Protocol} to coordinate the system.
As shown in Fig.~\ref{fig: 2 protocol}, the protocol contains three phases, i.e., the \emph{beaconing phase}, the \emph{positioning phase}, and the \emph{transmission phase}.
In the beaconing phase, the \acrshort{bs} broadcasts beacons for link construction. 
The users who want to communicate with the \acrshort{bs} reply acknowledgement frames to the \acrshort{bs}.
The \acrshort{bs} selects one of the users who have constructed an available link, responds to it, and proceeds to the next phase.

In the positioning phase, the selected user transmits signals to the \acrshort{bs} through \gls{numBand} bands sequentially, and the \acrshort{bs} receives the signals for $\gls{numFrame}$ frames in each band.
Specifically, in the frame~$q$~($q\in\{1,...,\gls{numFrame}\}$) of band $i$, the configuration of the \acrshort{rhs} is $\gls{codeVec_iq}$, and the combining vector is $\gls{sigVec_ijq}$ for sub-band $j$ of band $i$.
Besides, the combining vector of the \acrshort{bs} is bound by the maximum power \gls{userMaxPower}.
Denoting the received signals of the \acrshort{bs} in sub-band $j$ of band $i$ over the $\gls{numFrame}$ frames as vector $\bm y_{i,j}\in\mathbb C^{1\times \gls{numFrame}}$, the received signals in this phase can be arranged as matrix $\gls{recvSigMat}\in\mathbb C^{\gls{numBand}\times \gls{numSBand}\gls{numFrame}}$, whose block-wise elements are
$[\gls{recvSigMat}]_{i,j} = \bm y_{i,j}$.
At the end of this phase, the \acrshort{bs} uses a \emph{positioning function} to estimate the position of the user, and the result  is denoted by \gls{userPos_est}.
Moreover, despite incurring additional overhead in this phase, it should be noted that the user's transmission can also be used for sensing purposes:
A user can function as a monostatic radio-frequency sensor for the surrounding environment and motion~\cite{Chen2022ISACoT}. 
In this regard, this phase can be integrated into a higher-level ISAC network, and the overhead can be justified.

In the transmission phase, the \acrshort{bs} steers its beam to maximize the channel capacity towards \gls{userPos_est}. The \acrshort{bs} and the user start downlink/uplink transmission over one of the \gls{numBand} bands.

\subsection{Channel Model}
We establish the channel model by analyzing the equivalent base-band expression of the \acrshort{bs}'s received signals when a user at \gls{userPos} is transmitting.
Without loss of generality, we model the received signal in sub-band~$j$ of band~$i$ at frame~$q$.
Based on~\cite{goldsmith2005wireless}, for \acrshort{me}~$m$~($m\in \gls{valueSet_m}$), its received signal can be expressed as
\begin{align}
\label{equ: tau i j m}
& \tau_{i,j,m}^{(q)} = (\gls{losgain_ijm}(\gls{userPos}) + \gls{mpgain_ijmq}) \cdot x,
\end{align}
where $x$ indicates the \acrshort{tx} symbol of the user. 
Here, $\gls{losgain_ijm}$ is the gain of the \acrfull{los} path from the user at \gls{userPos} to the \acrshort{me} $m$, which can be expressed as 
\begin{align}
& \gls{losgain_ijm}(\gls{userPos}) = \frac{\gls{speedLight} \cdot g_i^{\elem} \cdot g_i^{\user} \cdot \exp(
-\iu \frac{2\pi f_{i,j}}{\gls{speedLight}} \cdot \|\gls{userPos} - \gls{posElem_m}\|_2
) }{4\pi f_{i,j} \cdot \|\gls{userPos} - \gls{posElem_m}\|_2}, \nonumber
\end{align}
where $\gls{speedLight}$ is the speed of light, 
$\iu$ is the imaginary unit, 
$g_i^{\user}$ and $g_i^{\elem}$ denote the gains of user's antenna and the \acrshort{me}, respectively, 
and $\gls{posElem_m}$ is the position of \acrshort{me}~$m$.

Besides, in~(\ref{equ: tau i j m}), $\gls{mpgain_ijmq}$ denotes the overall multipath gain.
Based on the multi-path gain model for \acrshort{ofdm} signals in rich-scattering environments~\cite{goldsmith2005wireless}, we can model $\gls{mpgain_ijmq}$ as a complex random variable following \acrfull{wss} Gaussian distribution.
Specifically, denoting $\gls{mpgainVec_ijq} = (h^{\mulpath,(q)}_{i,j,1},...,h^{\mulpath,(q)}_{i,j,\gls{numElem}})^\tran$ and with the help of~\cite{Barriac2006Space}, $\gls{mpgainVec_ijq}\sim \mathcal{CN}(0, \gls{covmat_i})$, where covariance matrix $\gls{covmat_i}$ can be obtained by the expectation of the outer product of \acrshort{rhs}'s array response $\gls{angularResponse_i}(\bm \theta)\in \mathbb C^{\gls{numElem}}$ over the angular domain, i.e., 
\beq
\label{equ: V i}
\bm V_i \!=\! \mathbb E \!\left( 
\gls{angularResponse_i}(\bm \theta) \gls{angularResponse_i}(\bm \theta)^{\hil} \right) 
\!=\!\oint \!\gls{angularResponse_i}(\bm \theta) \gls{angularResponse_i}(\bm \theta)^{\hil}\! P_{\mathrm{pap}}\!(\bm \theta) \!\dd \!\bm \theta, 
\eeq
where $[\gls{angularResponse_i}(\bm \theta)]_{m} = \exp(\iu \frac{2\pi \gls{cenFreq_i}}{\gls{speedLight}} (\gls{posElem_m} - \gls{posElem_1}) \cdot \hat{\bm n}(\bm \theta)) \cdot g^{\elem}_i$ with $\hat{\bm n}(\bm \theta)$ being the unit normal vector for $\bm \theta$, 
and $P_{\mathrm{pap}}(\bm \theta)$ is the \emph{power-angle profile} of the \acrshort{rhs}.
Based on~\cite{Barriac2006Space}, we model the covariance between the multi-path gains in the same band as 
\begin{align}
&\mathbb E\left[\bm h_{i, j_1}^{\mulpath, (q_1)}\big(\bm h_{i, j_2}^{\mulpath, (q_2)}\big)^{\hil}\right] = \gls{covCoefFuncFreq_i}(j_1,j_2)\cdot \gls{covCoefFuncTime_i}(q_1, q_2)\cdot \bm V_i,\nonumber\\
\label{equ: covCoefFuncFreq}
&\quad \gls{covCoefFuncFreq_i}(j_1,j_2) = (1+\iu 2\pi \gls{rmsValue_i} (f_{i,j_1} - f_{i, j_2}))^{-1},\\
\label{equ: covCoefFuncTime}
&\quad \gls{covCoefFuncTime_i}(q_1, q_2) = \mathcal J_0(2\pi \gls{dopplerFreq_i} (q_1 - q_2)\gls{durFrame}).
\end{align}

In~(\ref{equ: covCoefFuncFreq}), $\gls{rmsValue_i}$ indicates the \emph{root mean square~(rms) power delay spread} of band~$i$.
In~(\ref{equ: covCoefFuncTime}), $\mathcal J_0$ is the \emph{zeroth-order Bessel function of the first kind}, $\gls{dopplerFreq_i}$ indicates the maximum Doppler frequency, which can be calculated by $\gls{dopplerFreq_i} = \gls{userMaxSpeed} \gls{cenFreq_i} / \gls{speedLight}$, and \gls{durFrame} denotes the time duration of a frame.
We assume that different bands have larger spectral intervals than the coherence bandwidth of the channel, and thus the multi-path gains of different bands have zero covariance.

Then, based on~(\ref{equ: tau i j m}), the received signal at feed $k$~($k\in\{1,...,\gls{numFeed}\}$) is
\begin{align}
\label{equ: y i j k q}
& y_{i,j,k}^{(q)} = \sum_{m=1}^{\gls{numElem}} \tau_{i,j,m}^{(q)} \cdot \kappa(f_{i,j}, \bm p^{\mathrm{E}}_{m}, \bm p^{\mathrm{F}}_{k}),
\end{align}
where $\kappa(f_{i,j}, \bm p^{\mathrm{E}}_{m}, \bm p^{\mathrm{F}}_{k})$ is the onboard propagation gain.
Based on~\cite{Zhang2022Holographic}, it can be calculated as
\begin{align}
\label{equ: }
& \kappa(f, \bm p_{\mathrm{s}}, \bm p_{\mathrm{d}}) = \exp( -\iu \cdot\frac{2\pi n_{\mathrm{r}}f}{\gls{speedLight}} \cdot \|\bm p_{\mathrm{d}} - \bm p_{\mathrm{s}}\|_2 ),
\end{align}
where $n_{\mathrm{r}}$ is the relative refractive index of the \acrshort{rhs} board.

Combine the received signals at the $\gls{numFeed}$ feeds with weight $\gls{sigVec_ijq}$, and the received signal at the \acrshort{bs} can be modeled as
\begin{align}
\label{equ: y i j q}
& y_{i,j}^{(q)} =  \sum_{k=1}^{\gls{numFeed}} s_{i,j,k}^{(q)}y_{i,j,k}^{(q)} + e,
\end{align}
where $e \sim \mathcal{CN}(0, \gls{noisePower})$ is the thermal noise.
Given power spectral density of noise being \gls{noisePowerDensity}, noise power $\gls{noisePower} = \gls{noisePowerDensity}\gls{subBandwidth}$.

Finally, based on~(\ref{equ: tau i j m}),~(\ref{equ: y i j k q}), and~(\ref{equ: y i j q}), the received signals in the positioning phase can be formulated into matrix form, i.e., 
\begin{align}
\label{equ: recvSigVec_i}
& [\gls{recvSigMat}]_{i} =  \diag\big(\big(\bm  H_i^{\los}(\gls{userPos}) \otimes \bm 1_{\gls{numFrame}} + \bm H_i^{\mulpath}\big) \gls{holoMat_i}^\tran \big) x  + \bm e,
\end{align}
where the elements of the above matrices can be expressed as
\begin{align}
& [\bm H_i^{\los}(\gls{userPos})]_{j, m} \!=\! \gls{losgain_ijm}(\gls{userPos}),[\bm H_i^{\mulpath}]_{(q-1)\gls{numSBand}\!+\!j, m} \!=\! \gls{mpgain_ijmq}, \nonumber\\
\label{equ: matrix forms elements}
&[\bm T_i]_{j} = \bm C_i \odot \big( \bm S_{i,j} \bm B_{i,j} \big),~[\bm S_{i,j}]_{q} = \bm s_{i,j}^{(q)}, \\
&[\bm C_{i}]_{q} = \bm c_i^{(q)}, [\bm B_{i,j}]_{k,m} = \kappa(f_{i,j}, \bm p^{\mathrm{E}}_{m}, \bm p^{\mathrm{F}}_{k}). \nonumber
\end{align}
Based on channel reciprocity, when the \acrshort{bs} transmits, the received signals at the user can also be calculated by~(\ref{equ: recvSigVec_i}).

Furthermore, we model the channel capacity in the transmission phase of band $i$ given unit symbol.
Assuming that the \acrshort{bs} adopts combining vectors $\hat{\bm s}_{i,1}, ... \hat{\bm s}_{i, \gls{numSBand}}$ and \acrshort{rhs} configuration $\hat{\bm c}_i$, the channel capacity of the band $i$ can be calculated by 
\beq
\label{equ: capacity}
R_i = \sum_{j=1}^{\gls{numSBand}}\!W\!\log_2 (1 + \frac{\|\hat{\bm t}_{i,j}[\bm H_i^{\los}(\gls{userPos})]_{j}\|_2^2 + \hat{\bm t}_{i,j}\bm V_i \hat{\bm t}_{i,j}^{\hil}}{\gls{noisePower}}),
\eeq
where the received signal power of both the LoS and the multi-path channels are considered, and $\hat{\bm t}_{i,j} = \hat{\bm c}_i \odot \hat{\bm s}_{i,j}\bm B_{i,j}$.

\section{Problem Formulation}
\label{sec: problem formulation}

In this section, we formulate the \acrshort{and} beamforming optimization problem of the \gls{SysName}.

As in~\cite{Zohair2021Near}, we adopt the average \acrshort{crlb} in the \acrshort{roi} to evaluate the system's positioning performance, which indicates the lower-bound of the \acrfull{mse} of positioning.
Given the \acrshort{and} beamforming of the \acrshort{rhs}, the \acrshort{crlb} for the system to position a user at $\gls{userPos}$ can be derived by Proposition~\ref{prop: crlb expression}.

\begin{proposition}
\label{prop: crlb expression}
	The \acrshort{crlb} for \gls{SysName} to position a user at $\gls{userPos}\in \mathbb R^{1\times 3}$ can be calculated by
	\beq
\label{equ: general crlb expression}
\mathrm{CRLB}(\gls{userPos}) = \sum_{u=1}^3\big[
\bm I^{-1}_{\mathrm{FIM}} (\gls{userPos})
\big]_{uu}.
\eeq
Here, $\bm I_{\mathrm{FIM}}(\gls{userPos})\in\mathbb R^{3\times 3}$ is the \acrfull{fim} of $\gls{recvSigMat}$ w.r.t. $\gls{userPos}$, whose element can be calculated as:
\begin{align}
\label{equ: fim matrix}
& [\bm I_{\mathrm{FIM}} (\gls{userPos})]_{u,v} \!=\!  2\Re\Big(\sum_{i=1}^{\gls{numBand}}\! (\frac{\partial \hat{\bm y}_i}{\partial p_u^{\user}})^{\hil}\! \bm \varLambda_i^{-1} \!(\frac{\partial \hat{\bm y}_i}{\partial p_v^{\user}})\Big),\\
&   {\partial\hat{\bm y}_i \over \partial p_{u}^{\user}}= \diag\Big( (  \dot{\bm H}^{\los}_{i,u} \otimes \bm 1_{\gls{numFrame}} ) \gls{holoMat_i}^{\tran}\Big),~\dot{\bm H}^{\los}_{i,u}= {\partial\bm H_i^{\los}\over \partial p_u^{\user}},\\
&  \bm\varLambda_i = ( \bm K_{\mathrm{f},i} \!\otimes\! \bm J_{\gls{numFrame}})\!\odot\! ( \bm J_{\gls{numSBand}}\!\otimes\!\bm K_{\mathrm{t},i} ) \!\odot\! (\bm T_i \bm V_i \bm T_i^{\hil}) \!+\! \gls{noisePower}\bm I,\\
\label{equ: decorrelation matrices}
& [\bm K_{\mathrm{f},i}]_{j_1, j_2} =  \rho_{\mathrm{f},i}(j_1, j_2), ~[\bm K_{\mathrm{t},i}]_{q_1, q_2} =  \rho_{\mathrm{t},i}(q_1, q_2).
\end{align}
\end{proposition}
\begin{IEEEproof}
The \acrshort{fim} in~(\ref{equ: fim matrix}) can be obtained by substituting the channel model in~(\ref{equ: recvSigVec_i}) into the \acrshort{fim} Eq.~(6.55) in~\cite{Schreier2010Statistical}, and the \acrshort{crlb} can be derived based on Eq.~(27) in~\cite{Elzanaty2021Reconfigurable}.
\end{IEEEproof}

Then, the \acrshort{and} beamforming optimization problem of the system for \acrshort{crlb} minimization can be formulated as:
\begin{subequations}
\label{optprob: P1}
	\begin{align}
\text{(P1)}:~\min_{\gls{sigSet}, \gls{codeSet}}~&  \mathbb E_{\gls{userPos} \in \gls{userDistribu}}\big(\mathrm{CRLB}(\gls{userPos})\big),\\
\text{s.t.}~\qquad & \text{(\ref{equ: matrix forms elements})$\sim$(\ref{equ: decorrelation matrices}),}\\
& \sum_{j=1}^{\gls{numSBand}}(\gls{sigVec_ijq})^{\hil}\gls{sigVec_ijq} = \gls{userMaxPower},\\
& \bm 0\preceq \gls{codeMat_i} \preceq \bm 1.
\end{align}
\end{subequations}

The challenges in solving~(P1) lie in the following aspects.
\emph{Firstly}, due to the non-convex relationship between \gls{sigSet}, \gls{codeSet} and the \acrshort{crlb}, (P1) is non-convex and NP-hard.
\emph{Secondly}, the objective function (\ref{optprob: P1}a) is of high computational cost since it is calculated over the \acrshort{roi}.
\emph{Thirdly}, as multiple bands are used, the number of variables can be very large, i.e., $\gls{numBand}\gls{numFrame}(\gls{numSBand}\gls{numFeed}+\gls{numElem})$, which makes it hard for conventional algorithms to handle efficiently.
Therefore, it is necessary to design novel algorithm to solve (P1) efficiently.

\section{Alternating Analog and Digital Beamforming Optimization Algorithm}
\label{sec: alg design}

In this section, we propose an efficient algorithm named \emph{alternating \acrshort{and} beamforming optimization algorithm} to solve (P1).
\emph{Firstly}, to avoid the high complexity in finding the global optimum of a non-convex problem, we focus on gradient-based methods to search for a local optimum.
\emph{Secondly}, to reduce the complexity in calculating (\ref{optprob: P1}a), we adopt the Monte Carlo method to sample the \acrshort{roi}.
\emph{Thirdly}, to handle the large number of variables, we alternatingly optimize \acrshort{and} beamforming variables and derive the gradient formulas of the \acrshort{crlb} in a matrix form that is easy to calculate.

The \acrshort{roi} is sampled following distribution $\gls{userDistribu}$, and $\gls{numSamples}$ sampled positions are obtained and denoted by set $\gls{setSamples}$.
Then, (\ref{optprob: P1}a) is converted into:
\begin{align}
\label{equ: transformed obj func}
\min_{\gls{sigSet}, \gls{codeSet}}&  \sum_{\gls{userPos} \in \gls{setSamples}} \frac{\mathrm{CRLB}(\gls{userPos})}{\gls{numSamples}}.
\end{align}

We then solve the above problem by alternatingly optimizing $\{\gls{codeMat_i}\}_i$ and $\{\gls{sigMat_ij}\}_{i,j}$ in an iterative manner.
In the $\rho$-th iteration, $\{\gls{codeMat_i}\}_i$ is optimized given $\{\gls{sigMat_ij}\}_{i,j}^{(\rho-1)}$ by solving (sP1):
\begin{subequations}
\begin{align*}
\text{(sP1)}:~\min_{\gls{codeSet}}~ \sum_{\gls{userPos} \in \gls{setSamples}} {\mathrm{CRLB}(\gls{userPos})\over \gls{numSamples}}, \text{s.t.}~ \text{(\ref{equ: matrix forms elements})$\sim$(\ref{equ: decorrelation matrices}), (\ref{optprob: P1}d)}.
\end{align*}
\end{subequations}

To solve~(sP1) efficiently, we derive the gradient formulas of the objective average CRLB in Proposition~\ref{prop: grad of crlb with c}.
\begin{proposition}
\label{prop: grad of crlb with c}
The gradient of the average \acrshort{crlb} w.r.t. $\gls{codeMat_i}$ can be calculated as:\begin{align*}
& \sum_{\gls{userPos}\in \gls{setSamples}}\!{\partial \gls{crlb_userpos}\over \partial \gls{codeMat_i}} \!=\! - \hspace{-.4em}\sum_{\gls{userPos}\in \gls{setSamples}}\hspace{-.4em}\trace({\partial\gls{fimMat_userpos}\over \partial \gls{codeMat_i}}\gls{inv2fimMat_userpos}),\\
& \Big[{\partial\gls{fimMat_userpos}\over \partial \gls{codeMat_i}}\Big]_{u,v} = \bm A_{i,vu}^{\mathrm c} \!+\!  \bm A_{i,uv}^{\mathrm c} \!+\! \bm B_{i,uv}^{\mathrm c} \!+\! \bm B_{i,vu}^{\mathrm c} ,
\end{align*}
where $u,\!v \!\in\! \{1,\!2,\!3\}$. The detailed expressions of $\bm A_{i,uv}^{\mathrm c}$ and $\bm B_{i,uv}^{\mathrm c}$ can be found in Appendix~\ref{appx: 1}.
\end{proposition}
\begin{IEEEproof}
The gradient is obtained by calculating the partial derivatives of~(\ref{equ: general crlb expression}) w.r.t. \gls{codeMat_i},  based on~(36)$\sim$(40) in~\cite{matrix_cookbook}.
\end{IEEEproof}

Based on~Proposition~\ref{prop: grad of crlb with c}, (sP1) is solved by updating \gls{codeSet} along the inverse direction of the gradient.
As for the constraints in (sP1), constraints~(\ref{equ: matrix forms elements})$\sim$(\ref{equ: decorrelation matrices}) are implicitly contained in the calculation of gradients.
Besides, constraint~(\ref{optprob: P1}d) can be handled by the \emph{interior point approach}, where a barrier function is used to restrict $\gls{codeSet}$ within its feasible region~\cite{Byrd2000Trust}.
Moreover, we adopt \emph{\acrfull{cg} method} to calculate the update by solving a quadratic approximation of (sP1) within a local region of current $\gls{codeSet}$~\cite{Byrd2000Trust}.
The result of solving (sP1) in the $\rho$-th iteration is denoted by $\{\bm C_{i}\}_i^{(\rho)}$.

Given $\{\bm C_{i}\}_i^{(\rho)}$, $\{\gls{sigMat_ij}\}_{i,j}$ is optimized by solving (sP2):
\begin{subequations}
\begin{align}
\text{(sP2)}:~\min_{\gls{sigSet}}~ \hspace{-0.3em}\sum_{\gls{userPos} \in \gls{setSamples}}\! {\mathrm{CRLB}(\gls{userPos})\over \gls{numSamples}},~
\text{s.t.}~\text{(\ref{equ: matrix forms elements})$\sim$(\ref{equ: decorrelation matrices}), (\ref{optprob: P1}c)}. \nonumber
\end{align}
\end{subequations}

Similar with solving~(sP1), we adopt gradient-based method to solve~(sP2), where the gradient is derived in Proposition~\ref{prop: grad of crlb with s}.
 \begin{proposition}
\label{prop: grad of crlb with s}
The gradient of the average \acrshort{crlb} w.r.t. $\gls{sigMat_ij}$ can be calculated as
\begin{align*}
& \sum_{\gls{userPos}\in \gls{setSamples}}\hspace{-1em}{\partial \gls{crlb_userpos}\over \partial \gls{sigMat_ij}} \!=\! - \hspace{-0.0em}\sum_{\gls{userPos}\in \gls{setSamples}}\hspace{-0.0em}\trace({\partial\gls{fimMat_userpos}\over \partial \gls{sigMat_ij}}\!\gls{inv2fimMat_userpos}),\\
& [{\partial\gls{fimMat_userpos}\over \partial \gls{sigMat_ij}}]_{u,v} =  \bm A_{ij,vu}^{\mathrm s}\!+\!  \bm A_{ij,uv}^{\mathrm s} \!+\! \bm B_{ij,uv}^{\mathrm s} \!+\! \bm B_{ij,vu}^{\mathrm s},
\end{align*}
where $u,\!v \!\in\! \{1,\!2,\!3\}$.
The detailed expressions of $\bm A_{ij,uv}^{\mathrm s}$ and $\bm B_{ij,uv}^{\mathrm s}$ can be found in Appendix~\ref{appx: 1}.
\end{proposition}
\begin{IEEEproof}
The gradient is obtained by calculating partial derivatives of~(\ref{equ: general crlb expression}) w.r.t. \gls{sigMat_ij},  based on~(36)$\sim$(40) in~\cite{matrix_cookbook}.
\end{IEEEproof}

We solve (sP2) also by gradually updating each $\gls{sigSet}$ along the inverse direction of its gradient, and the interior point approach is used to handle constraint~(\ref{optprob: P1}c) with \acrshort{cg} method utilized to calculate the update.
The result of solving~(sP2) in the $\rho$-th iteration is denoted by $\{\bm S_{i,j}\}_{i,j}^{(\rho)}$.

Within each iteration, the number of update steps for $\{\gls{codeMat_i}\}_i$ and $\{\gls{sigMat_ij}\}_{i,j}$ are denoted by \gls{numUpdate}.
The iteration process terminates when the results are not changed by the last update or when the number of iterations exceeds $\gls{numMaxIteration}$.
The complete algorithm is summarized as Algorithm~1.

\begin{algorithm}[!t]
\small
  \caption{Alternating Analog and Digital Beamforming Optimization Algorithm}
\label{alg: summary algorithm}
\begin{algorithmic} [1]
\State Sample $\gls{numSamples}$ positions following $\gls{userDistribu}$ and obtain $\gls{setSamples}$.
\State Set initial points by $\{\gls{sigMat_ij}\}_{i,j}^{(0)} = \{\gls{sigMat_ij}|\gls{sigVec_ijq}=\bm 1\cdot \sqrt{\gls{userMaxPower}}/\gls{numFeed}\}$ and $\{\gls{codeMat_i}\}_i^{(0)}$ with elements randomly chosen from $[0,1]$.
\For{$\rho = 1, ..., \gls{numMaxIteration}$}
\State Given $\{\gls{sigMat_ij}\}_{i,j}^{(\rho-1)}$ and $\{\gls{codeMat_i}\}_i^{(\rho-1)}$, solve (sP1) to obtain $\{\gls{codeMat_i}\}_i^{(\rho)}$ using the gradient formula in Proposition~\ref{prop: grad of crlb with c}.
\State Given $\{\gls{codeMat_i}\}_i^{(\rho)}$ and $\{\gls{sigMat_ij}\}_{i,j}^{(\rho-1)}$, solve (sP2) to obtain $\{\gls{sigMat_ij}\}_{i,j}^{(\rho)}$ using the gradient formula in Proposition~\ref{prop: grad of crlb with s}.
\State \textbf{If} $\{\gls{sigMat_ij}\}_{i,j}^{(\rho-1)}\!=\!\{\gls{sigMat_ij}\}_{i,j}^{(\rho)}$ and $\{\gls{codeMat_i}\}_i^{(\rho-1)}\!=\!\{\gls{codeMat_i}\}_i^{(\rho)}$: \textbf{break}
\EndFor
\State \Return $\{\gls{sigMat_ij}\}_{i,j}^{*}=\{\gls{sigMat_ij}\}_{i,j}^{(\rho)}$ and $\{\gls{codeMat_i}\}_i^{*}=\{\gls{codeMat_i}\}_i^{(\rho)}$.
\end{algorithmic}
\end{algorithm}

\section{Simulation Results}
\label{sec: simulation result}

In this section, we present the simulation setup and the key results.
We establish a 3D coordinate system with its origin at the center of \acrshort{rhs}, its x-axis along the perpendicular direction of \acrshort{rhs}, and its z-axis pointing vertically upward.
The \acrshort{roi} is a cuboid region centering at $(10,0,0)$~m and has dimensions $(10,10,2)$~m.
The distribution of users' positions, i.e. $\gls{userDistribu}$, is a 3D uniform distribution within the \acrshort{roi}.
The gains of each \acrshort{me} and user's antenna are $g_i^{\elem}=g_{i}^{\user}=1$, while it is worth noticing that the proposed system and algorithm can adapt to \acrshort{me}s with other patterns of gain.

Moreover, the \acrshort{rhs} board in the simulation is made of FR-4, which is a typical dielectric material used for printed circuit boards and has $n_{\mathrm{r}}=2.1$.
The \gls{numBand} bands of the system are centered at $(2+0.5i)$~GHz with a  $500$~MHz interval, and the average wavelength of the center frequencies is denoted by $\lambda_{\mathrm{avr}}$.
The interval between adjacent \acrshort{me}s is set to be $0.3\lambda_{\mathrm{avr}}$.
Besides, based on~\cite{Barriac2006Space}, $P_{\mathrm{pap}}(\bm \theta)$ in~(\ref{equ: V i}) is modeled as a Laplacian function with zero mean and angular spread $10^{\circ}$ in both azimuth and elevation, scaled by the average LoS power within the \acrshort{roi}.
The other parameters are listed in Table~\ref{table: parameter}.

\begin{table}
\vspace{-1em}
\centering
\caption{Simulation Parameters}
\label{table: parameter}
\begin{small}
\begin{tabular}{|l|l|l|l|}
\hline
\textbf{Parameter}     & \textbf{Value}                                & \textbf{Parameter}            & \textbf{Value}   \\ \hline\hline
\gls{numBand}           & $2$                                                  & \gls{numElem}                   & $10\times 10$   \\ \hline
\gls{numFeed}           & $3$                                                  & \gls{rmsValue_i}                 & $0.5~\mu$s       \\ \hline
\gls{numSBand}         & $8$                                                 & \gls{userMaxSpeed}          & $20$ km/h         \\ \hline
\gls{durFrame}           & $4~\mu$s                                      & \gls{noisePowerDensity}   & $-174$ dBm/Hz \\ \hline
\gls{numFrame}         &   $4$                                                & \gls{userMaxPower}           & $1$ mW            \\ \hline
\gls{subBandwidth}   & $125$ kHz                                       & \gls{numSamples}             & $1200$              \\ \hline
\gls{numUpdate}        &   $50$                                             & \gls{numMaxIteration}       & $4$                   \\ \hline
\end{tabular}
\end{small}
\vspace{-0.1em}
\end{table}

\emph{Firstly}, we show the effectiveness of the proposed algorithm in terms of \acrshort{crlb} minimization and the computational efficiency.
We compare it with two benchmark algorithms: 
\begin{itemize}[leftmargin=*]
\item \textbf{Direct Gradient Descent}: Problem (P1) is solved by using gradient descent directly without alternating variables. Gradients are calculated based on the Propositions~\ref{prop: grad of crlb with c} and~\ref{prop: grad of crlb with s}. For a fair comparison, the total number of update steps is set to $2\times \gls{numMaxIteration} \times \gls{numUpdate}$, and the interior point approach with \acrshort{cg} method is also adopted. 
\item \textbf{Genetic Algorithm}: The genetic algorithm is employed as in~\cite{Nguyen2020Reconfigurable}, which minimizes the \acrshort{crlb} by optimizing \gls{sigSet} and \gls{codeSet} jointly. The standard Matlab implementation is adopted with the maximum number of generations set to $10$.
\end{itemize}

\begin{figure}[!t]
\centerline{ \includegraphics[width=1\linewidth]{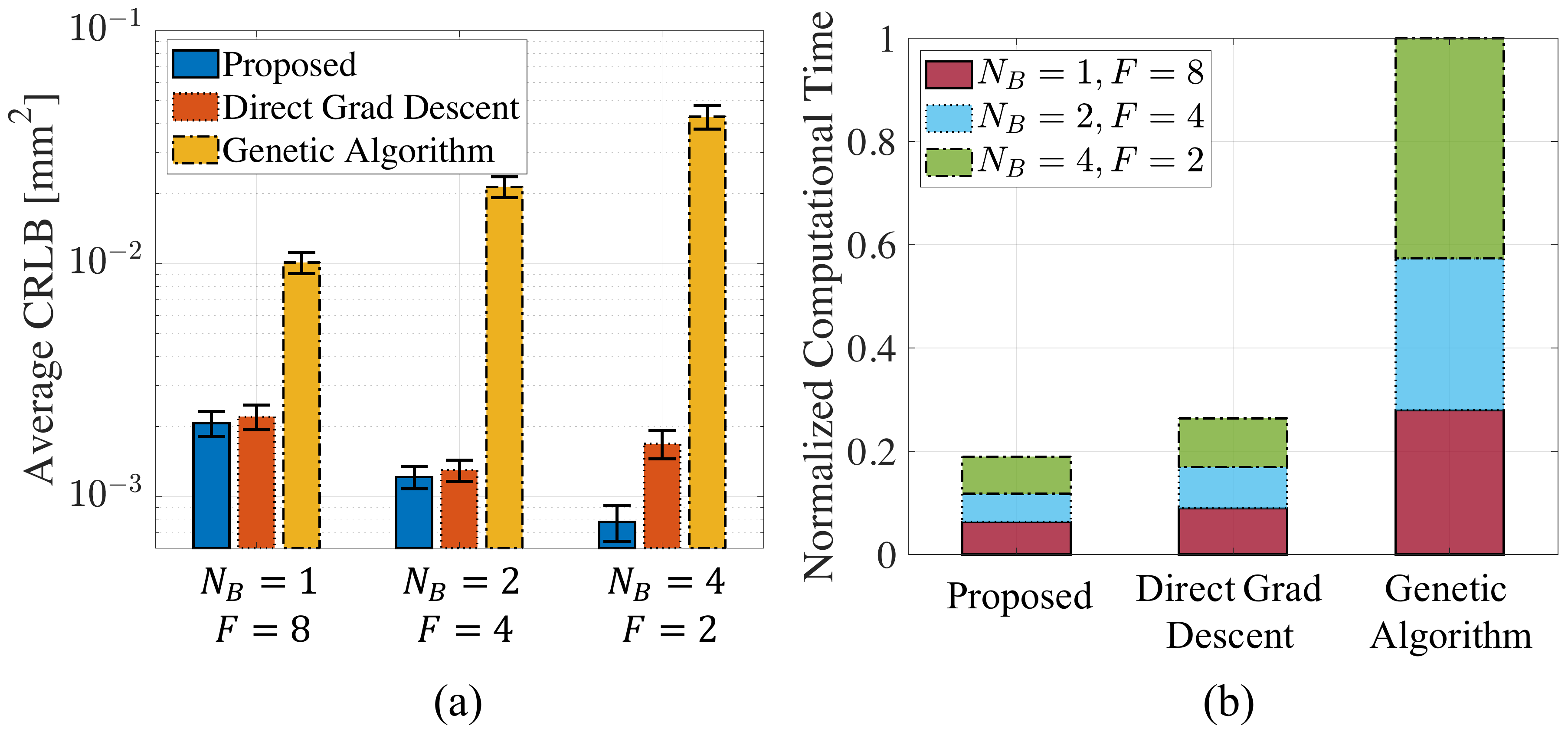} }
\vspace{-0.5em}
\caption{Average optimization results of (a) CRLB and (b) computational time of the proposed and benchmark algorithms. \gls{numSamples} is set to $10$ ensure fast output, and $30$ independent trails are performed to reduce the impact of sampling randomness. The length of the ``I''-shaped bars in (a) represents the standard deviations of the results.}
\label{fig: simul res 1}
\end{figure}

Figs.~\ref{fig: simul res 1} (a) and (b) show the comparison results of the resulting \acrshort{crlb} and the normalized computational time for the proposed and benchmark algorithms, respectively.
It can be observed that the proposed algorithm results in lower \acrshort{crlb} values and takes less computational time compared with the benchmark algorithms.
Fig.~\ref{fig: simul res 1} (a) also shows that the resulting \acrshort{crlb} of the proposed algorithm decreases with \gls{numBand} given fixed $\gls{numBand}\gls{numFrame}$, which proves the effectiveness of using \acrshort{mb} technique.
Moreover, the \acrshort{crlb} gap between the proposed algorithm and the benchmarks increases with \gls{numBand}.
This implies that the advantage of the proposed algorithm becomes more pronounced as \gls{numBand} increases.

\emph{Secondly}, we show the effectiveness of the optimized \acrshort{and} beamforming in terms of its resulting positioning precision and its benefit for the communication capacity.
We compare the optimized \acrshort{and} beamforming with two benchmark beamforming used in~\cite{Zohair2021Near}:
\begin{itemize}[leftmargin=*]
\item \textbf{Directional}: The \acrshort{and} beamforming is designed to generate focused beams scanning the \acrshort{roi} during the frames in the positioning phase.
\item \textbf{Random}: Elements of \gls{codeSet} follow a uniform distribution in range $[0,1]$, and elements of \gls{sigSet} take uniform values that satisfy the power constraint.
\end{itemize}

\begin{figure}[!t]
\centerline{ \includegraphics[width=1\linewidth]{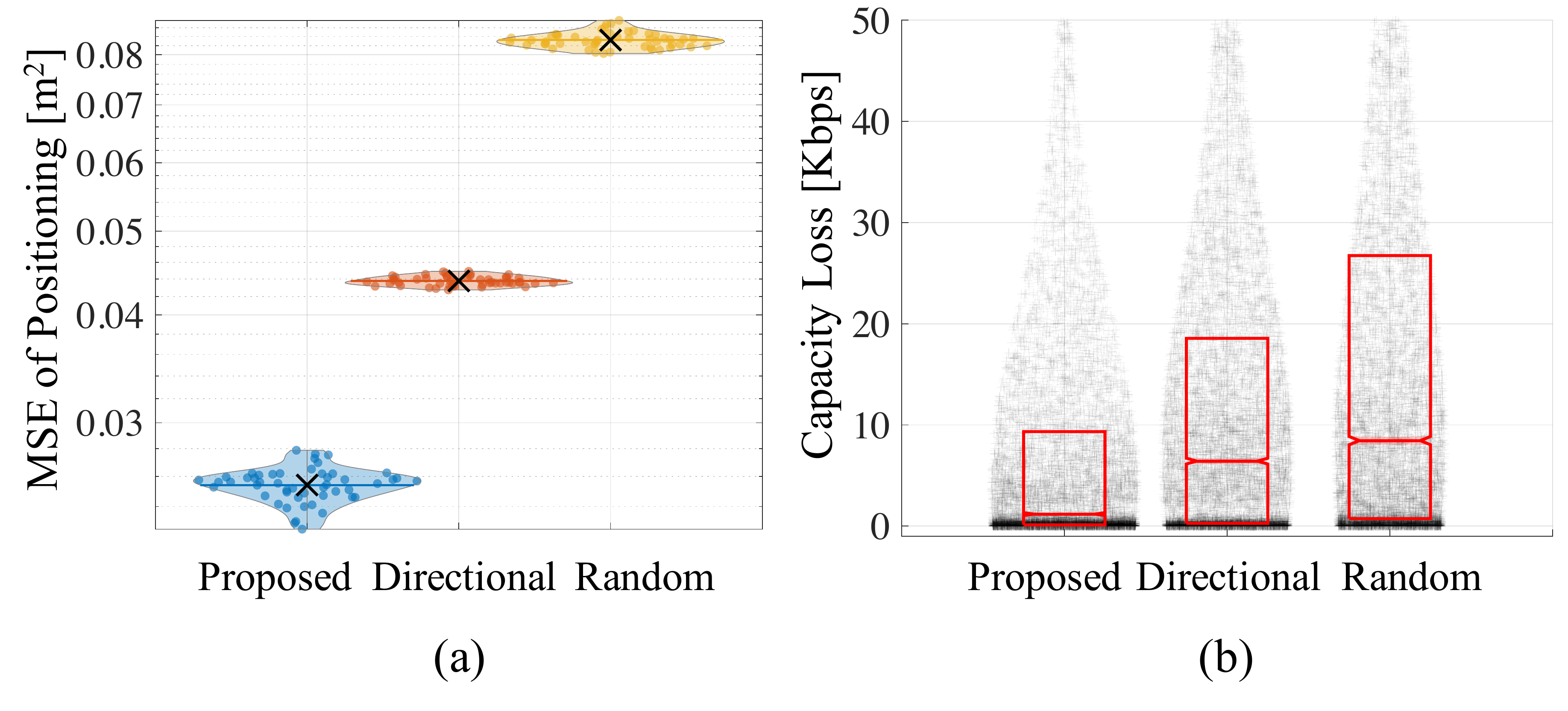} }
\vspace{-0.1em}
\caption{(a) \gls{mse} of positioning and (b) communication capacity loss due to positioning error given different \acrshort{and} beamforming. In (a), $50$ independent training are conducted for the neural network. Each dot represents the result of a single trial, the filled regions indicate the overall distribution of the results, and ``x'' represents the mean value. In (b), the best training result of each case is used, and $1000$ random positions are sampled. The translucent background figures indicate the patterns of the capacity loss of individual samples. The red boxplots show the distribution of the results, with the bottom, middle, and top lines indicating the $25$th, $50$th, and $75$th percentiles, respectively.}
\label{fig: simul res 2}
\vspace{-0.5em}
\end{figure}

Figs.~\ref{fig: simul res 2} (a) and (b) show the violin plot and the box plot comparing the performance of different beamformings in terms of the \acrshort{mse} of positioning and the communication capacity loss due to positioning errors, respectively of $256$ nodes.
Specifically, in Fig.~\ref{fig: simul res 2}~(a), the \acrshort{mse} of positioning is evaluated by the training result of a neural network with a single hidden-layer.
The neural networks in different cases are trained by datasets each containing $10^5$ pairs of simulated received signals and position labels obtained by the system with different beamforming.
During the training, the division between training, validation, and test data follows ratio $7\!:\!1.5\!:\!1.5$.
It can be observed in Fig.~\ref{fig: simul res 2}~(a) that, on average, by using the optimized \acrshort{and} beamforming, the system reduces \acrshort{mse} of positioning by $42\%$ and $69\%$, compared to the Directional and Random beamforming benchmarks.

Besides, in Fig.~\ref{fig: simul res 2}~(b), the communication capacity loss is evaluated by the difference between the capacity of band $1$ when the \acrshort{bs} steers its beam towards user's true position and that when the beam is steered towards the estimated position.
It can be observed that the optimized \acrshort{and} beamforming results in $82\%$ and $86\%$ reduction in median levels of capacity loss, compared with Directional and Random benchmarks, respectively.
Moreover, the interquartile range, i.e., the range between 25\% and 75\% levels, of the capacity loss is also significantly reduced by the proposed algorithm, which are $50\%$ and $64\%$ lower than those of the benchmarks.

\section{Conclusion}
\label{sec: conclu}
In this paper, we have proposed the \gls{SysName}.
We have designed the \gls{Protocol}, established the channel model, and  analyzed the \acrshort{crlb} of the system.
Then, we have designed an \acrshort{and} beamforming optimization algorithm for the \acrshort{crlb} minimization, which handles the large number of beamforming variables efficiently.
Simulation results have shown that firstly, the proposed algorithm achieves lower average \acrshort{crlb} with less computational time compared with benchmark algorithms.
Secondly, the \acrshort{and} beamforming obtained by the proposed algorithm results in $42\%$ lower \acrshort{mse} of positioning, which leads to $82\%$ less communication capacity loss, compared with the benchmark directional beamforming.

\vspace{-.3em}
\begin{appendices}
\section{}
\label{appx: 1}
\begin{small}
With the help of~\cite{matrix_cookbook}, $\bm A_{i,uv}^{\mathrm c}$, $\bm B^{\mathrm c}_{i,uv}$, $\bm A_{ij,uv}^{\mathrm s}$, and $\bm B^{\mathrm s}_{ij,uv}$ ($\forall u,v\in\{1,2,3\}$) can be derived as
\begin{align*}
\bm A_{i,uv}^{\mathrm c}  = & 2\Re\big( \sum_{j=1}^{\gls{numSBand}}  ([\bar{\bm \zeta}_{i,u}]_{(j-1)\gls{numFrame}+1:j\gls{numFrame}}\otimes \bm 1_{\gls{numElem}}^\tran) \\
& \odot [\dot{\bm H}^{\los}_{i,v}]_{(j-1)\gls{numFrame}+1:j\gls{numFrame}}
\odot (\bm S_{i,j}\bm B_{i,j})\big),\\
\bm B_{i,uv}^{\mathrm c} =&   -2\Re\big(\sum_{j}^{\gls{numSBand}} 
\big([\bar{\bm \zeta}_{i,u}]_{(j-1)\gls{numFrame}+1:j\gls{numFrame}}\otimes \bm 1_{\gls{numElem}}^\tran \odot (\bm S_{i,j} \bm B_{i,j})\big) \nonumber \\
& \odot \gls{reshapeFunc}_{\gls{numFrame}\!\times\!\gls{numElem}}\big(
[\bm K_{\mathrm{ft},i}]_{(j-1)\gls{numFrame}+1:j\gls{numFrame}}
[\odot]\big((\bm V_i\bm T_i^{\hil})\otimes \bm 1_{\gls{numFrame}}\big) \bm\zeta_{i,v}
\big)\big),\\
\bm A_{ij,uv}^{\mathrm s} =& 2\Re\big([\bar{\bm \zeta}_{i,u}]_{(j-1)\gls{numFrame}+1:\gls{numFrame}}\otimes \bm 1_{\gls{numFeed}}^\tran \\
&\hspace{-1.5em}\odot \! \gls{reshapeFunc}_{\gls{numFrame}\!\times\!\gls{numFeed}}\big(\big(([\dot{\bm H}_{i,v}^{\los}]_{(j-1)\gls{numFrame}+1:j\gls{numFrame}} \!\odot\! \bm C_i)[\odot] (\bm B_{i,j} \!\otimes\! \bm 1_{\gls{numFrame}}) \big) \bm 1_{\gls{numElem}}\big)\big),\\
\bm  B_{ij,uv}^{\mathrm s} = & - 2\Re\big(([\bar{\bm \zeta}_{i,u}]_{(j-1)\gls{numFrame}+1:j\gls{numFrame}} \!\otimes\! \bm 1_{\gls{numFeed}}^\tran)\\
&\hspace{-4em} \odot \!\gls{reshapeFunc}_{\gls{numFrame}\!\times\!\gls{numFeed}}\big(\!\big(
[\bm K_{\mathrm{ft},i}]_{(j-1)\gls{numFrame}+1:j\gls{numFrame}} [\odot] \big(
(\bm C_i [\odot] \bm B_{i,j} \otimes \bm 1_{\gls{numFrame}})(\bm V_i\bm T_i^{\hil})
\big)\!\big)
\bm\zeta_{i,v}
\big)\!\big),
\vspace{0.5em}
\end{align*}
where $\bm\zeta_{i,u} =\bm\varLambda_i^{-1}(\dot{\bm H}^{\los}_{i,u}\odot \bm T_i) \bm 1_{\gls{numElem}}\in\mathbb C^{\gls{numSBand}\gls{numFrame}\times 1}$, 
operator $[\odot]$ denotes the \emph{penetrating face product}, 
$\bm K_{\mathrm{ft},i}= ( \bm K_{\mathrm{f},i} \!\otimes\! \bm J_{\gls{numFrame}})\!\odot\! ( \bm J_{\gls{numSBand}}\!\otimes\!\bm K_{\mathrm{t},i} )$,
and $\gls{reshapeFunc}_{\gls{numFrame}\!\times\!\gls{numFeed}}(\cdot)$ reshapes the argument into a $\gls{numFrame}\!\times\!\gls{numFeed}$ matrix.
\end{small}
 
 \end{appendices}

\bibliographystyle{IEEEtran}
\bibliography{ms.bib}
\end{document}